\documentclass[preprint]{aastex62}            

\usepackage{amsmath}
\usepackage{bm}
\usepackage{multirow}
\usepackage{xcolor}
\usepackage{graphicx}

\shorttitle{Dicarbon formation in collisions of two carbon atoms}
\begin{document}

\title{Dicarbon formation in collisions of two carbon atoms}

\author[0000-0002-3883-9501]{James F. Babb}
\affil{Institute for Theoretical Atomic, Molecular, and Optical Physics,\\
Center for Astrophysics \textbar \ Harvard \& Smithsonian, 60 Garden St., Cambridge, MA 02138}
\email{jbabb@cfa.harvard.edu}

\author[0000-0002-4359-1408]{R. T. Smyth}
\affil{Institute for Theoretical Atomic, Molecular, and Optical Physics,\\
Center for Astrophysics \textbar \ Harvard \& Smithsonian, 60 Garden St., Cambridge, MA 02138}
\affil{Centre for Theoretical Atomic, Molecular, and Optical Physics, School of Mathematics \& Physics \\ 
Queen's University of Belfast, Belfast BT7 1NN, Northern Ireland, UK}
\email{rsmyth41@qub.ac.uk}

\author[0000-0002-5917-0763]{B. M. McLaughlin}
\affil{Institute for Theoretical Atomic, Molecular, and Optical Physics,\\
Center for Astrophysics \textbar \ Harvard \& Smithsonian, 60 Garden St., Cambridge, MA 02138}
\affil{Centre for Theoretical Atomic, Molecular, and Optical Physics, School of Mathematics \& Physics \\ 
Queen's University of Belfast, Belfast BT7 1NN, Northern Ireland, UK}
\email{bmclaughlin899@btinternet.com}

\date{Accepted 2019 March 14}

\label{firstpage}

\begin{abstract}
Radiative association cross sections and rates are computed, using a quantum approach, for the formation 
of C$_2$ molecules (dicarbon) during the collision of two ground state C($^3$P) atoms. 
We find that transitions originating in the C$\;^1\Pi_g$, d$\;^3\Pi_g$, and 1$\;^5\Pi_u$ states
are the main contributors to the process.
The results 
are compared and contrasted with previous results obtained from a semi-classical approximation. 
New \textit{ab initio} potential curves and transition dipole moment functions 
have been obtained for the present work using the multi-reference configuration interaction 
approach with the Davidson correction (MRCI+Q) and aug-cc-pCV5Z basis sets,
substantially increasing the available molecular data on dicarbon.
Applications of the current computations to various astrophysical environments
and laboratory studies are briefly discussed focusing on these rates.
\end{abstract}


\section{Introduction}\label{sec:introduction}

Dicarbon (C$_2$) was first observed spectroscopically in flames and arcs and continues
to be a useful diagnostic there, and in carbon plasmas for other laboratory and industrial applications \citep{nemes11}. 
The molecule has been observed in a host of extraterrestrial sources such
as comets, carbon stars, protoplanetary nebulae, and molecular clouds.

Interstellar dicarbon has been
detected at optical wavelengths in diffuse \citep{ChaLut78,ewine93,gredel01} 
and translucent \citep{ewine89,iglesias-groth11} molecular clouds. 
The optical detection of C$_2$ in comets is an element of their 
classification into ``typical'' and ``depleted''  \citep{ahearn95,cochran12}.
Dicarbon is present in solar \citep{lambert78} and model stellar atmospheres, including
the pioneering work of \citet{tsuji64} and \citet{lord65}, 
and seen, for example, in solar optical \citep{grevesse73} and infrared spectra \citep{brault82}
and in infrared spectra of 
carbon-rich giant stars \citep{goebel83,loidl01}.  
The presence of the Swan bands of dicarbon (optical wavelengths) is an important element
in the classification scheme of carbon stars~\citep{keenan93,green13}.
The mechanisms of formation of dicarbon vary, depending on the operative
chemistries: For example, in diffuse molecular clouds 
dissociative recombination of $\mbox{CH}_2^+$ leads to $\mbox{C}_2$~\citep{black77,federman89},
while in comets a chemistry starting with photodissociation
of $\mathrm{C}_2\mathrm{H}_2$ or $\mathrm{C}_2\mathrm{H}$ may be operative~\citep{jackson76}.

Of particular interest for the present work is the formation of carbonaceous dust
in the ejecta of core-collapse supernovae, where the formation of dicarbon through
radiative association enters chemical models~\citep{liu92,cherchneff09,clayton18,sluder18}
and is an initial step in models of formation of larger carbon clusters
through condensation \citep{clayton99,clayton01,clayton13} or nucleation \citep{lazzati16}.
[Later, in Sec.~\ref{sec:discussion}, we discuss in more
detail laboratory experiments on carbon vapors generated by laser radiation.
We note at this point that evidence of associative collisions of two ground state carbon atoms was found by \citet{monchicourt91} in light emission
from laser-induced expansion of carbon vapor from a graphite rod.]
\begin{figure}
    \includegraphics[width=\textwidth]{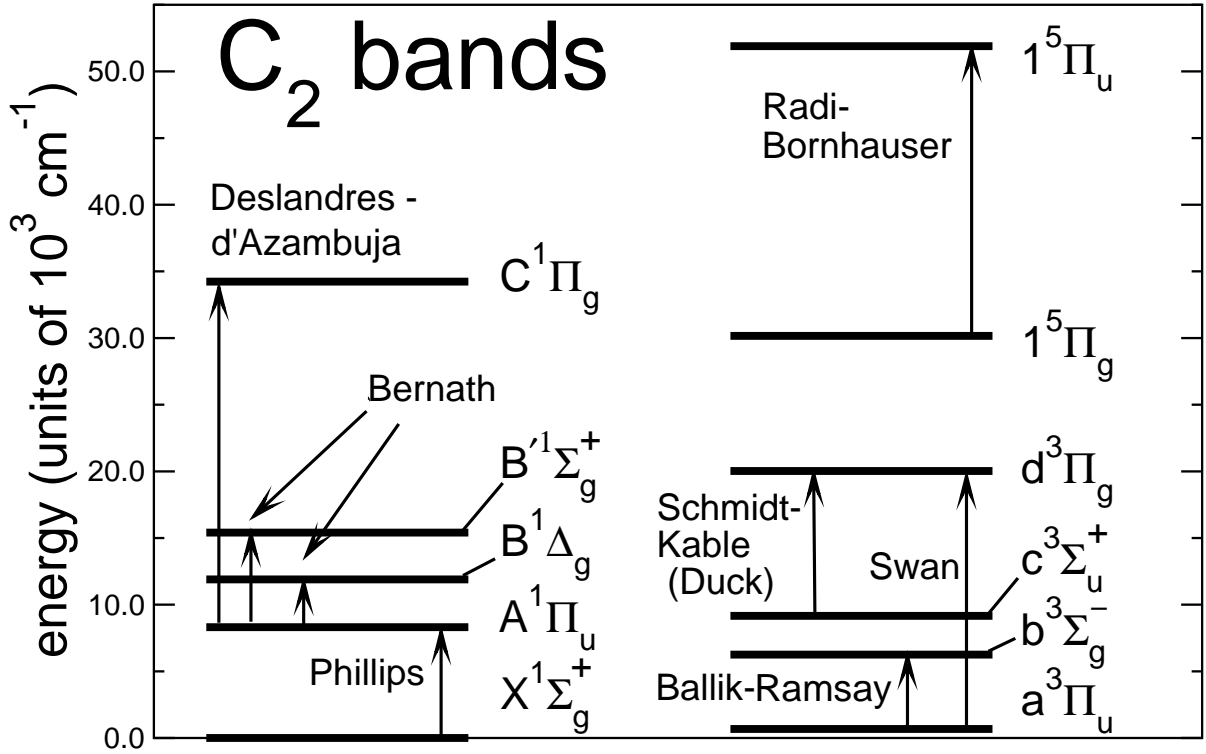}
    \caption{Experimentally observed C$_2$ band systems connecting eleven singlet, triplet, and quintet electronic states 
                    dissociating to ground state carbon atoms; schematic illustration of electronic state term energies $T_e$ in $\mbox{cm}^{-1}$
                    calculated in the present work [after \citet{messerle67,tanabashi07,bornhauser15,macrae16,furtenbacher16}].}
        \label{fig:bands}
\end{figure}
Fig.~\ref{fig:bands} illustrates a sample of the experimentally 
observed bands~\citep{tanabashi07,bornhauser15,macrae16,furtenbacher16} connecting eleven singlet, triplet, and
quintet states of the C$_2$ molecule that 
contribute to the overall radiative association rate coefficient for this molecule.
In the ejecta of SN1987A and other core-collapse supernovae, CO and SiO were detected, see~\citep{cherchneff11,sarangi18}, through
fundamental $(\Delta\nu=1)$ bands\footnote{CO was also detected in the first overtone band $(\Delta\nu =2)$. In SN 1987A individual
rotational lines of CO and of SiO were detected at late epoch, see~\protect\citet{abellan17,sarangi18}.}
of ground molecular electronic states
allowing observational tests of molecular
formation models~\citep{liu92,liu95,cherchneff11,sarangi13,rho18}---dicarbon, however, 
lacks a permanent electric dipole moment and analogous vibrational transitions
(fundamental bands)
do not exist, making reliable theoretical predictions
of rate coefficients imperative.
Moreover, recent three-dimensional mapping of CO and SiO in the SN 1987A core ejecta
with the Atacama Large Millimeter/submillimeter Array (ALMA) shows a clumpy mixed structure 
calling for improvements beyond one-dimension in hydrodynamical and chemical modeling of molecular
formation~\citep{abellan17}; a reliable description of dicarbide formation might improve
such future calculations.
Finally, understanding the origins of cosmic dust and the roles
played by supernovae in contributing to extragalactic dust
depends on progress in modeling dust formation \citep{sarangi18,sluder18}.

The radiative association (RA) rate was originally estimated to have a rate coefficient 
$k_{C_2} \approx 1 \times 10^{-17}$ cm$^3$/s \citep{prasad80,millar91}
for theoretical models of interstellar clouds and subsequent
semi-classical calculations \citep{andreazza97} found
comparable values with a weak temperature dependence increasing
from $3.07 \times 10^{-18}$ cm$^3$/s at 300~K to $1.65 \times 10^{-17}$ cm$^3$/s at $14,700$~K.
In recent studies using a quantum approach 
on systems such as SiP \citep{golubev13}, SiO \citep{forrey16,forrey17}  and CS \citep{patillo18,forrey18}, 
it was found that the semi-classical calculations
\citep{andreazza97,andreazza06} underestimated
the cross sections and rates, particularly at low temperatures.

In the present study we obtain results from a quantum approach to 
estimate the cross sections and rate coefficients for C$_2$ formation by radiative association
using new highly accurate ab initio molecular data 
for the potential energy curves (PEC's) and transition dipole moments (TDM's) coupling 
the states of interest.  Results from our present quantum approach are compared 
with the previous semi-classical results of \citet{andreazza97}
and conclusions are drawn.

The layout of this paper is as follows. An overview of how the molecular data is obtained for 
our dynamical calculations  is presented in \autoref{sec:theory}. In \autoref{sec:results}, 
a brief overview of the radiative association cross section 
and rates are outlined.  The computed radiative association cross sections, and rates  are presented  in 
\autoref{sec:results} and are compared with the previous semi-classical 
work of \citet{andreazza97} in \autoref{sec:discussion}. Finally in \autoref{sec:conclusions} conclusions are drawn from our work. 
Atomic units are used throughout unless otherwise specified.

\section{Theory and Calculations}\label{sec:theory}

\subsection{Potential Curves and Transition Dipole Moments}\label{sec:pecstdms}

In a similar manner to our recent all electron molecular structure and resulting dynamical studies on diatomic systems such as: SiO 
\citep{forrey16,forrey17}, CS \citep{patillo18,forrey18}, HeC$^+$ \citep{babb17a},  SH$^+$ \citep{shen15,mcmillan16},
CH$^+$ \citep{babb17b}, and HeAr$^+$ \citep{babb18a},
the potential energy curves (PECs) and transition dipole moments (TDMs) for the eighteen singlet, triplet and quintet electronic states
are calculated within an MRCI+Q approximation for the approach of ground state carbon atoms. That is, we use a 
state-averaged-multi-configuration-self-consistent-field (SA-MCSCF) approach, followed by multi-reference 
configuration interaction (MRCI) calculations together with the Davidson correction (MRCI+Q) \citep{Helgaker2000}. 
The SA-MCSCF method is used as the reference wave function for the MRCI calculations. 
Low-lying singlet, triplet and quintet electronic states and the transition dipole matrix elements connecting these 
molecular states are calculated and used in the present dynamical calculations for the radiative association process.
The literature on the molecular properties of dicarbon is extensive;
sources with comprehensive bibliographies include \citet{martin92,nemes11,zhang11,boschen14,macrae16,furtenbacher16,yurchenko18,varandas18}.
  
Potential energy curves and transition dipole moments as a function of internuclear distance 
$R$ are calculated starting from a bond separation of $R = 1.5$ Bohr extending out to $R=20$ Bohr. 
The basis sets used in the present work are the augmented correlation consistent 
 polarized core valence quintuplet [aug-cc-pcV5Z (ACV5Z)] Gaussian basis sets. 
The use of such large basis sets is well known to recover 98\% of 
the electron correlation effects in molecular structure calculations \citep{Helgaker2000}. 
All the PEC and TDM calculations for the C$_2$ molecule were performed with the 
quantum chemistry program package \textsc{molpro} 2015.1 \citep{molpro2015}, 
running on parallel architectures.
	
For molecules with degenerate symmetry, an Abelian subgroup is required to be used in 
\textsc{molpro}. So for a diatomic molecule like C$_2$ with D$_{{\infty}h}$ symmetry, 
it will be substituted by D$_{2h}$ symmetry with the 
order of irreducible representations  being 
($A_g$, $B_{3g}$, $B_{2g}$, $B_{1g}$, $B_{1u}$, $B_{2u}$, $B_{3u}$,  $A_u$). 
When symmetry is reduced from D$_{{\infty}h}$ to D$_{2h}$, \citep{herz50} 
the correlating relationships are 
$\sigma_g \rightarrow a_g$, 
$\sigma_u \rightarrow a_u$, 
$\pi_g \rightarrow$ ($b_{2g}$, $b_{3g}$) , 
$\pi_u \rightarrow$ ($b_{2u}$, $b_{3u}$) , 
$\delta_g \rightarrow$ ($a_g$, $b_{1g}$), 
 and 
$\delta_u \rightarrow$ ($a_u$, $b_{1u}$). 

In order to take account of short-range interactions, 
we employed the non-relativistic state-averaged 
complete active-space-self-consistent-field (SA-CASSCF)/MRCI method 
available within the \textsc{molpro} \citep{molpro2012,molpro2015} quantum chemistry suite of codes.  

For the C$_2$ molecule,  molecular orbitals (MOs) are put into the active space, 
including (3$a_g$,  1$b_{3u}$, 1$b_{2u}$, 0$b_{1g}$, 3$b_{1u}$, 1$b_{2g}$, 1$b_{3g}$, 0$a_u$),  symmetry MOs.
The molecular orbitals for the MRCI procedure were obtained using the 
SA-MCSF method, for singlet and triplet spin symmetries,
 we carried out the averaging processes on 
the two lowest states of the symmetries; ($A_g$, $B_{3u}$, $B_{1g}$, $B_{1u}$)
and the lowest states of the symmetries; ($B_{2u}$, $B_{3g}$, $B_{2g}$ and $A_u$).  
A similar approach was also used for the quintet states.  
This approach provides an accurate representation of the singlet, triplet and quintet  
states of interest as the molecule dissociated.

At  bond separations beyond $R = 14$ Bohr, the PECs are smoothly 
fitted to functions of the form
\begin{equation}\label{eq:longrange}
    V(R) = \frac{C_5}{R^5}-\frac{C_6}{R^6}\,,
\end{equation}
where for the particular electronic
state, $C_5$ is the quadrupole-quadrupole electrostatic interaction~\citep{Knipp38,chang67} and $C_6$ is the 
dipole-dipole dispersion 
(van der Waals) coefficient  (we use atomic units unless otherwise specified).
For $R < 1.5$~Bohr,  short-range interaction potentials 
of the form $V(R) = A \exp(-BR)+C$ are fitted to the \textit{ab initio} potential curves. 
Estimates of the values of the quadrupole-quadrupole
coefficients $C_5$ were given by \citet{Knipp38},
and by \citet{boggio-pasqua00}
(for  singlet and triplet $\Sigma$, $\Pi$ and $\Delta$ electronic states, which suffices to determine those for quintet states by symmetry).
The long range dispersion coefficient $C_6$ (averaged over the possible
fine structure levels of two carbon atoms)
was calculated to be $40.9 \pm 4.4$ using many-body perturbation theory by \citet{MilKel72}
and estimated to be 46.29 using the London formula by \citet{chang67}.
In fitting the long-range form Eq.~(\ref{eq:longrange}) to the calculated potential energy data, we adjusted
the values of $C_5$ and $C_6$, as necessary to match the 
slopes of the potential energy curves. The adopted values
are given in Table~\ref{table:states}.
We began with
estimates of $C_5$ from \citet{Knipp38} and 
\citet{boggio-pasqua00} and limited our adjustment of $C_6$ to either the value of \citet{MilKel72} or that of \citet{chang67}.
For the B$^{\prime}{}^1\Sigma_g^+$ state, there is a barrier in
the potential energy curve (0.0086~eV or 69~$\textrm{cm}^{-1}$ at $R=7.12$),
reflected in the positive value of $C_5$ which fit the data.
This is in good accord with a value found by \citet{varandas08}, 43~$\textrm{cm}^{-1}$ at $R=8.37$,
in an extensive study of diabatic representations of the X$^1\Sigma_g^+$  and B$^{\prime}{}^1\Sigma_g^+$ states.
(This barrier energy is too low to appreciably affect the dynamics calculations presented below.)

\begin{deluxetable}{c@{$\,$}lDDDD}[ht!]
	\centering
    \tablecaption{The eighteen C$_2$, singlet, triplet and quintet electronic states
    formed during the collision of two ground state carbon atoms.
    In column 2 the values of the term energies $T_e$ are listed for
    the potential energies fit to Eq.~(\ref{eq:longrange}) relative to the minimum
    of the $\textrm{X}^1\Sigma_g^+$ potential energy.
    For each electronic state used in the present work, we list in atomic units values
    of $C_5$ from \citet{boggio-pasqua00} and the adopted values of $C_5$
    and $C_6$ entering Eq.~(\ref{eq:longrange}).
    The states 1$^1\Sigma_u^-$, 1$^5\Delta_g$, and 1$^5\Sigma_u^-$ were not fitted in the present work,
    so we list the values of $T_e$ as calculated \textit{ab initio} for 1$^1\Sigma_u^-$ and 1$^5\Delta_g$,
    while 1$^5\Sigma_u^-$ is repulsive.
    \label{table:states}}
    \tablehead{
    \twocolhead{State} & 
    \twocolhead{$T_e$ ($\textrm{cm}^{-1}$)\tablenotemark{a}} &
    \twocolhead{$C_5$\tablenotemark{b}} &
    \twocolhead{$C_5$\tablenotemark{c}} &
    \twocolhead{$C_6$\tablenotemark{d}} 
    }
    \decimals
    \startdata
    X&$^1\Sigma_g^+$ 				& 0.0 			& 21.81	 		& 5	 		& 40.9	 \\ 
    A&$^1\Pi_u$ 				 	& 8312.4 		& 0.			& 0.	 	& 46.29	 \\ 
    B&$^1\Delta_g$ 		     		& 11895. 		& 3.635		 	& 3.635	 	& 46.29	 \\ 
    B$^{\prime}$&$^1\Sigma_g^+$ 	& 15205. 		& 0. 			& 16.	 	& 40.29	 \\ 
    C&$^1\Pi_g$ 				 	& 34231. 		& $-$14.54		& $-$13.49	& 40.9	 \\ 
    1&$^1\Sigma_u^-$        	 	& 39500.		& 0.			& .	 		& .	 	 \\
    \hline
    a&$^3\Pi_u$ 					& 678.74  		& $-$14.54		& $-$14.54	& 46.29  \\ 
    b&$^3\Sigma_g^-$ 			 	& 6254.9		& 0. 			& 0. 		& 40.9	 \\ 
    c&$^3\Sigma_u^+$ 			 	& 9157.5  		& 21.81  		& 5.		& 40.9	 \\ 	
    d&$^3\Pi_g$ 					& 20031.  		& 0. 			& 0.		& 46.29	 \\ 	
    2&$^3\Sigma_u^+$ 				& 40042. 		& 0.  			& 0.		& 40.9	  \\ 	
    1&$^3\Delta_u$ 					& 41889.  		& 3.635			& 8. 	 	& 46.29	 \\ 
    \hline
    1&$^5\Pi_g$ 					& 30165. 		& $-$14.54	 	& $-$13.49	& 40.9	 \\ 
    1&$^5\Sigma_g^+$ 				& 40197.		& 0. 			& 0. 		& 40.9	 \\ 
    1&$^5\Pi_u$ 					& 51897.  		& 0.		 	& 0.	 	& 46.29  \\ 	
    2&$^5\Sigma_g^+$ 				& 64088.  		& 21.81  		& 12.		& 40.29	 \\ 	
    1&$^5\Delta_g$          	 	& 49900.		& 3.635	 		& .	 		& .	 	 \\
    1&$^5\Sigma_u^-$            	& . 			& 0.  			& .			& .	 	 \\
    \enddata
    \tablenotetext{a}{Present calculations.} 
    \tablenotetext{b}{Singlet and triplet values from \citet{boggio-pasqua00}, quintet values by symmetry. See text for details.}
    \tablenotetext{c}{Actual value used. See text for details.}
    \tablenotetext{d}{Estimated from \citet{MilKel72} or \citet{chang67}. See text for details.}
\end{deluxetable}

\begin{figure}
    \centering
        \gridline{\fig{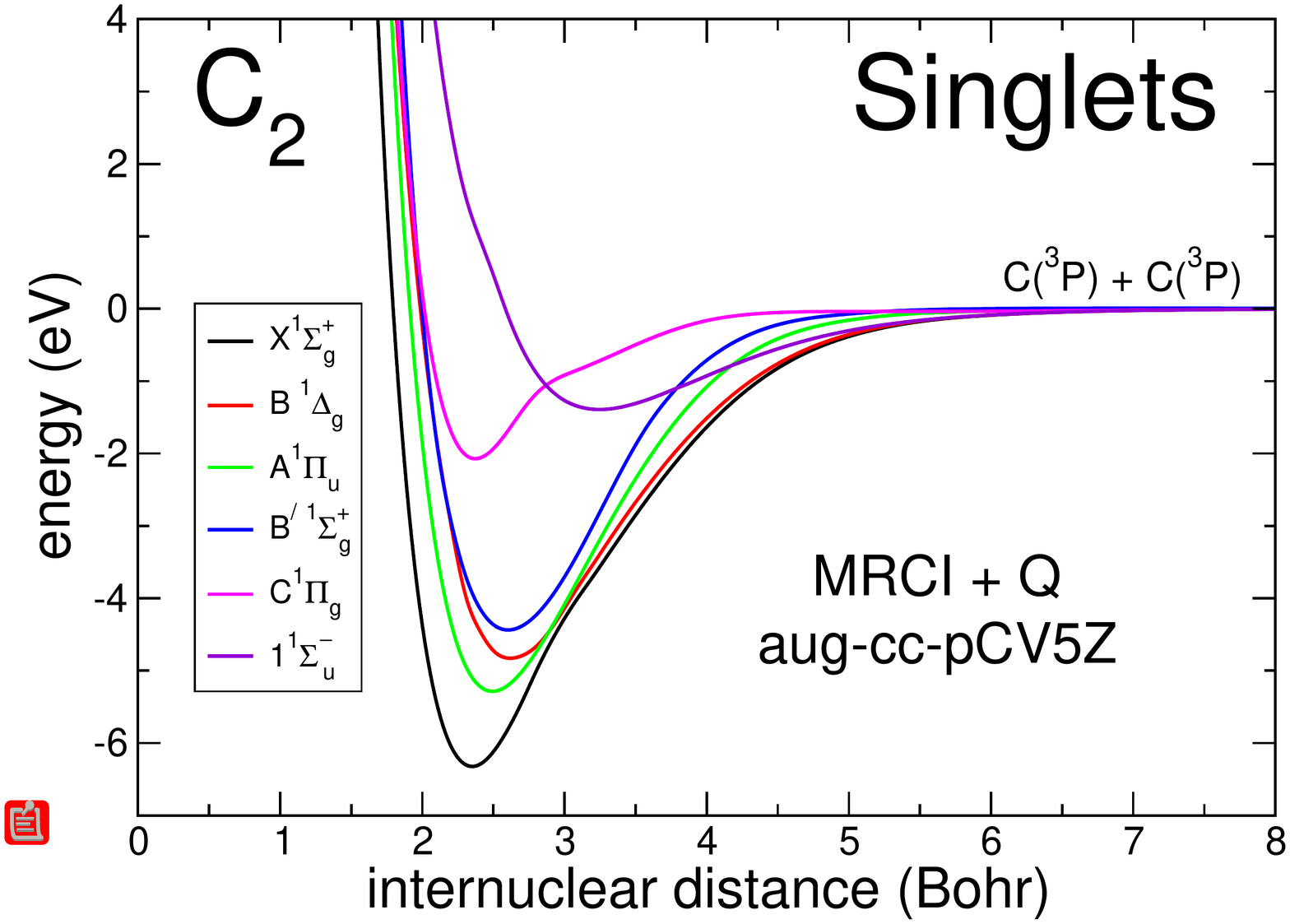}{8.9cm}{(a)}
                        \fig{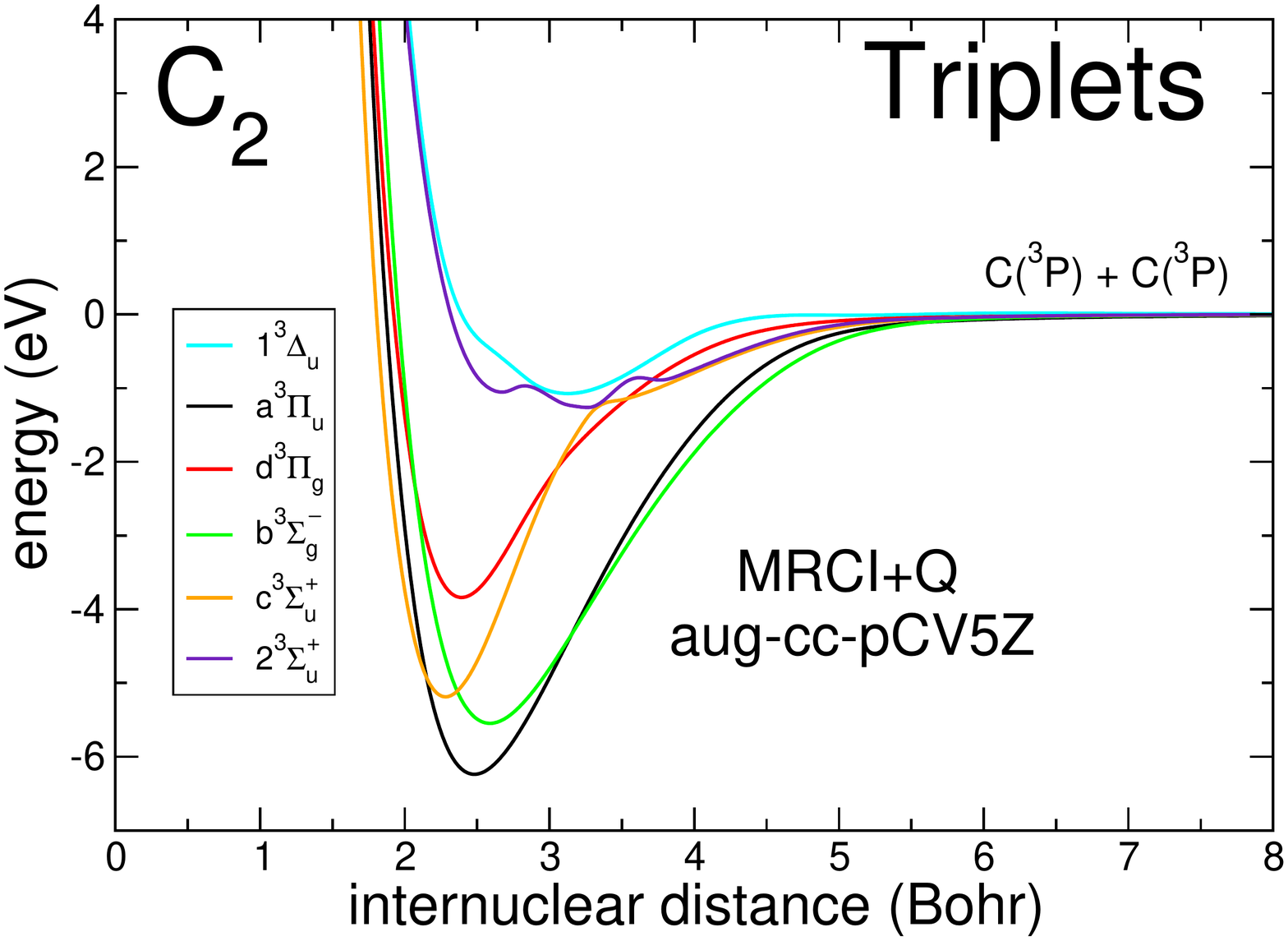}{8.9cm}{(b)}}
                         \fig{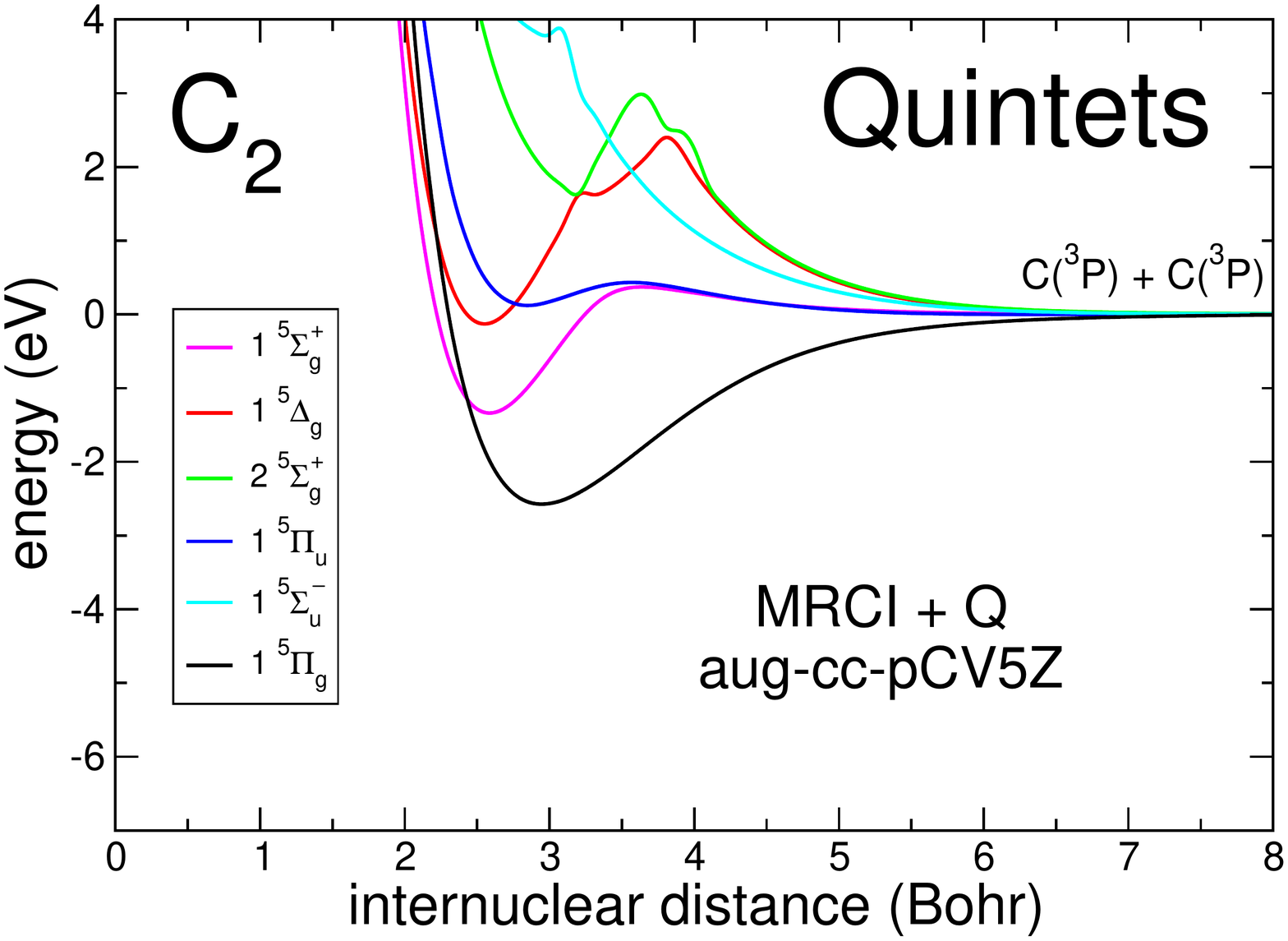}{8.9cm}{(c)}
       \caption{Potential energy curves (eV), as a function of internuclear distance (Bohr) for 
                      C$_2$ molecular states dissociating to ground state carbon atoms, (a) singlet, (b) triplet, and (c) quintet states.
                       Results were obtained using the quantum chemistry package \textsc{molpro} and 
                       aug-cc-pCV5Z basis for each atom.}
    \label{fig:pots}
\end{figure}

As a consequence of fitting the potentials to Eq.~(\ref{eq:longrange}),
the calculated potentials (as output from \textsc{molpro})
were shifted.
In Table~\ref{table:states} we list the final values of $T_e$ (in $\textrm{cm}^{-1})$ relative
to the minimum of the $\textrm{X}^1\Sigma_g^+$  potential energy curve
and the term energies are plotted schematically in Fig.~\ref{fig:bands}
for experimentally observed bands.
Our calculated values may be compared with the recent experimental fits in Table~4, column~3, of \citet{furtenbacher16},
and we agree to within 100~$\textrm{cm}^{-1}$ for
the a$^3\Pi_u$, A$^1\Pi_u$, c$^3\Sigma_u^+$, and B$^1\Delta_g$ states, and to within 205~$\textrm{cm}^{-1}$ for the b$^3\Sigma_g^-$ 
and B$^{\prime}{}^1\Sigma_g^+$ states.
For the 1$^5\Pi_g$ state, our value of $T_e$ is within 9~$\textrm{cm}^{-1}$
of the \textit{ab initio} value given by \citet{schmidt11} (expressed with respect
to the minimum of the a$^3\Pi_u$ curve and calculated using the aug-cc-pV5Z basis sets).
Our calculated 1$^5\Pi_u$ state supports
a shallow well in agreement with previous calculations \citep{bruna01,bornhauser15,Visser2019}, deepening
the shelf-like form of the curve found in the earlier calculations of  \citet{kirby79}. 
The accuracy of the present potential
energy curves is very satisfactory for the purposes of the present study.
The potential curves for C$_2$ singlet, triplet and quintet states are shown in Fig.~\ref{fig:pots}. 
Structures due to nonadiabatic couplings are apparent in the c$^3\Sigma_u^+$ and  2$^3\Sigma_u^+$
curves due to their mutual interaction and interactions with higher states (see Fig.~3 of \citet{kirby79}).
As we will show below, the calculated cross sections involving the c$^3\Sigma_u^+$ and  2$^3\Sigma_u^+$
states are several orders of magnitude smaller than those from the  leading transitions
contributing to the total radiative association cross sections.
Similar structures in the quintet states~\citep{bruna01} occur at energies at about 1~eV 
corresponding to kinetic temperatures above the range of the present study.
Thus, we ignore nonadiabatic couplings and expect that they will have at most a minor effect on the net total
cross sections for radiative association.

The TDMs for the C$_2$  molecule are similarly extended to long- and short-range internuclear distances. 
For $R > 14$ a functional fit of the form $D(R) = a \exp(-bR) + c$ is applied, 
while in the short range $R < 1.5$ a quadratic fit of the form $D(R) = a' R^2 + b' R + c'$ is adopted. 
\begin{figure}
    \centering
    \gridline{\fig{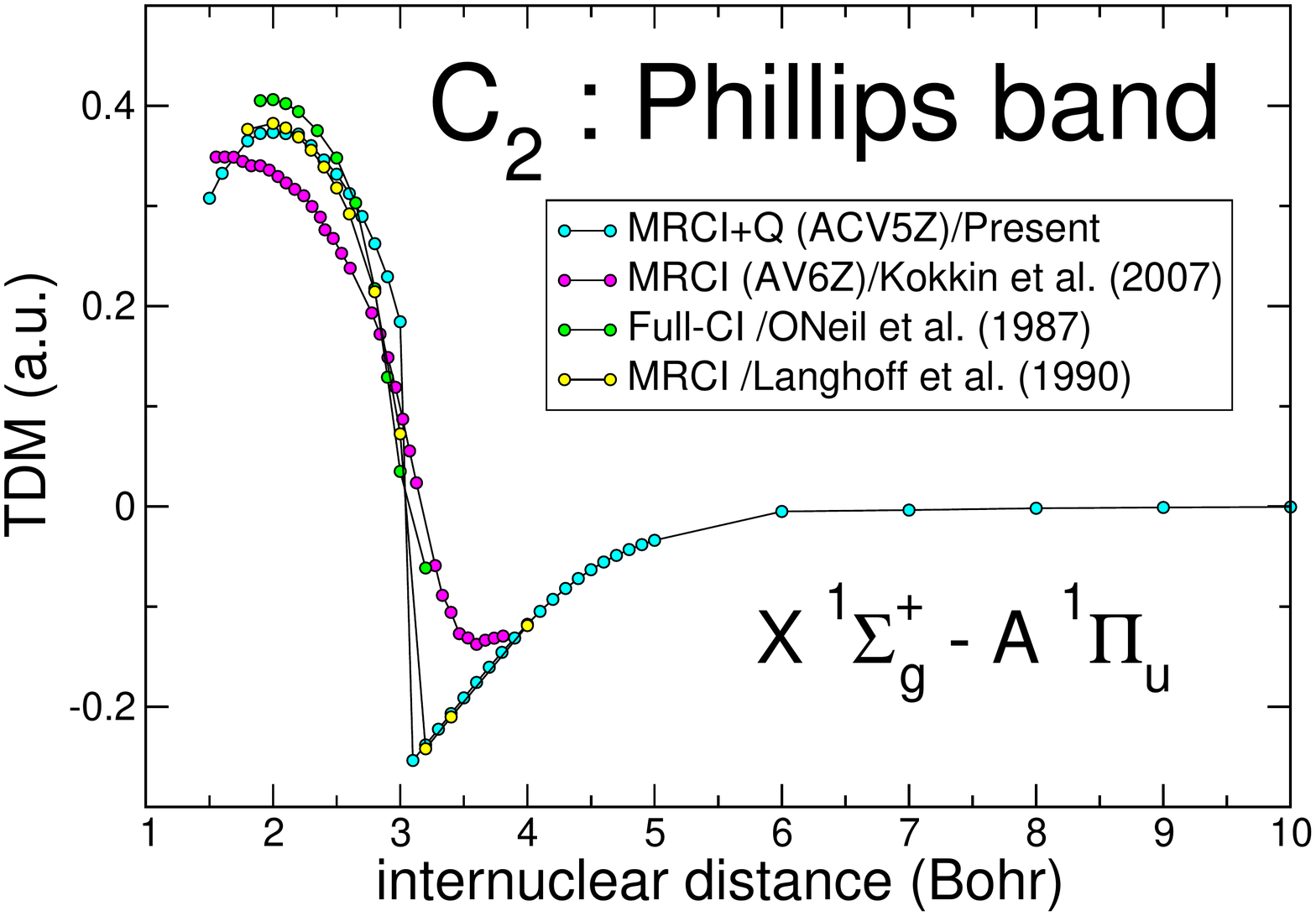}{8.9cm}{(a)}
             \fig{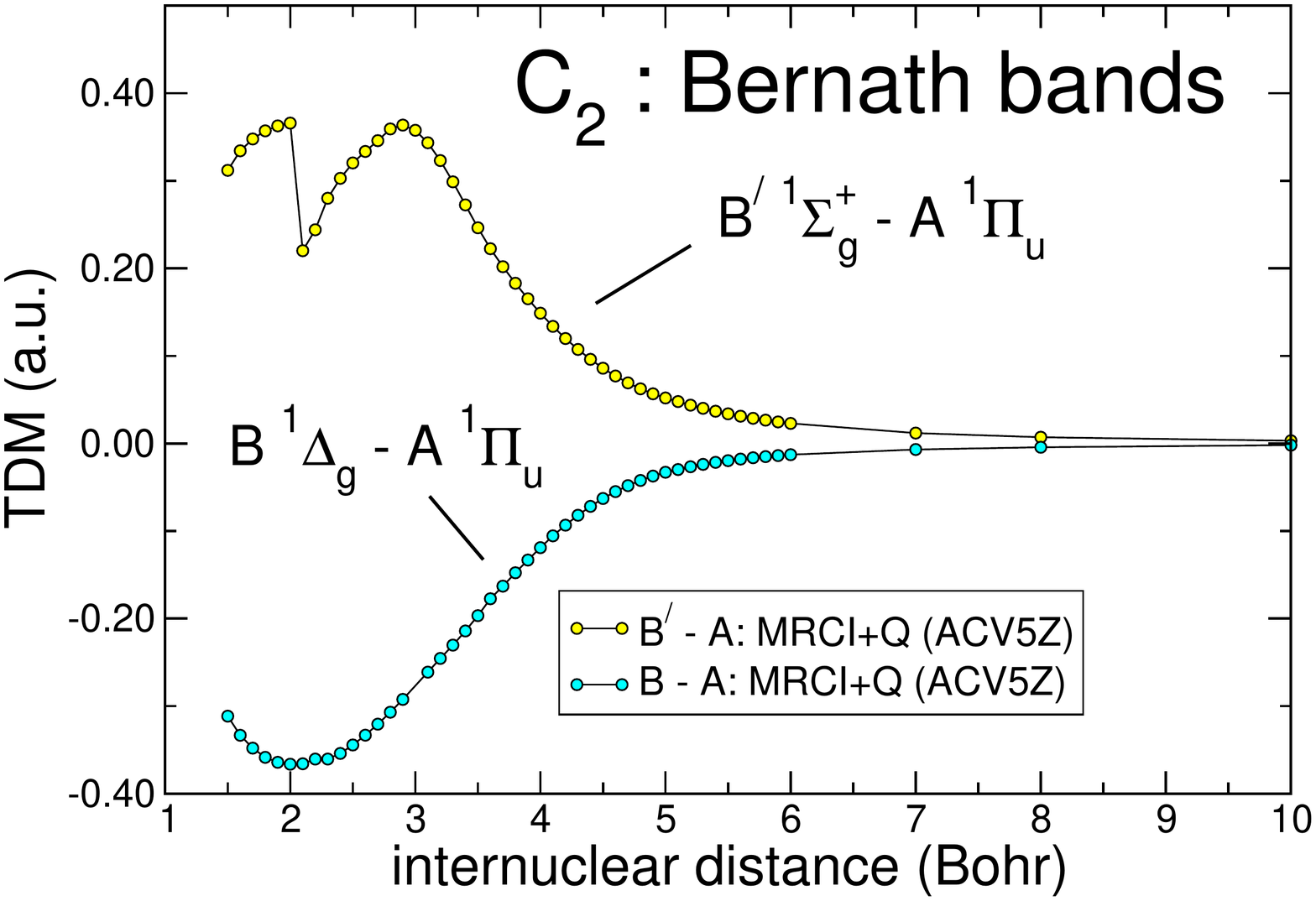}{8.9cm}{(b)}
             }
    \fig{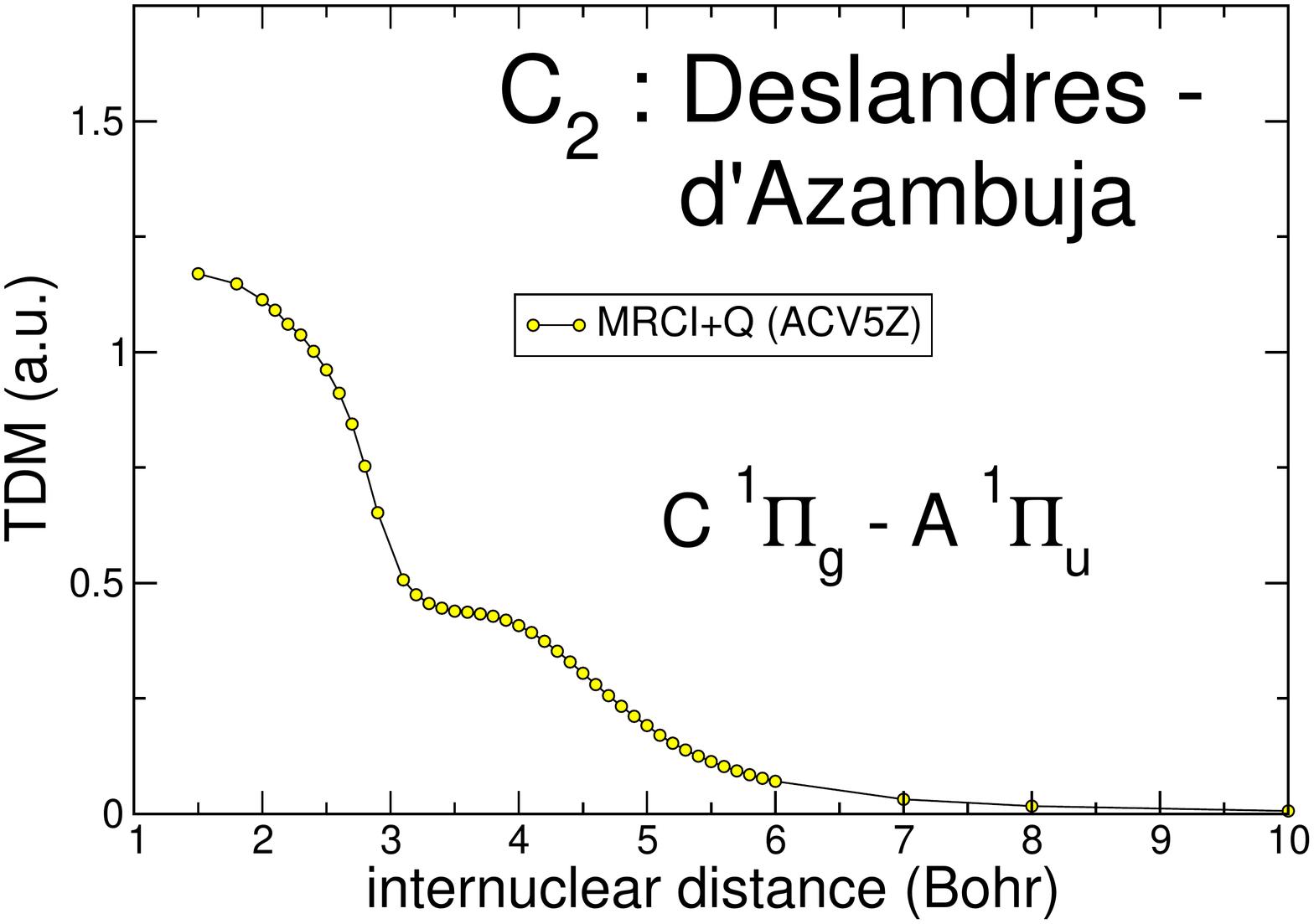}{8.9cm}{(c)}
    \caption{Transition dipole moments (TDMs) for singlet transitions in C$_2$ in atomic units, 
                    (a) Phillips band (A$^1\Pi_g$-X$^1\Sigma_g^+$), 
                    (b) Bernath bands (B$^1\Delta_g$-A$^1\Pi_u$ and B$^{\prime}\Sigma_g^+$-A$^1\Pi_u$)  and 
                    (c) Deslandres-d'Azambuja band (C$^1\Pi_g$-A$^1\Pi_u$).
                    For the Phillips band we compare the present MRCI+Q work with results from \citet{oneil87,langhoff90}
                    and \citet{kokkin07}.}
    \label{fig:tdmsing}
\end{figure}
\begin{figure}
    \centering
        \gridline{\fig{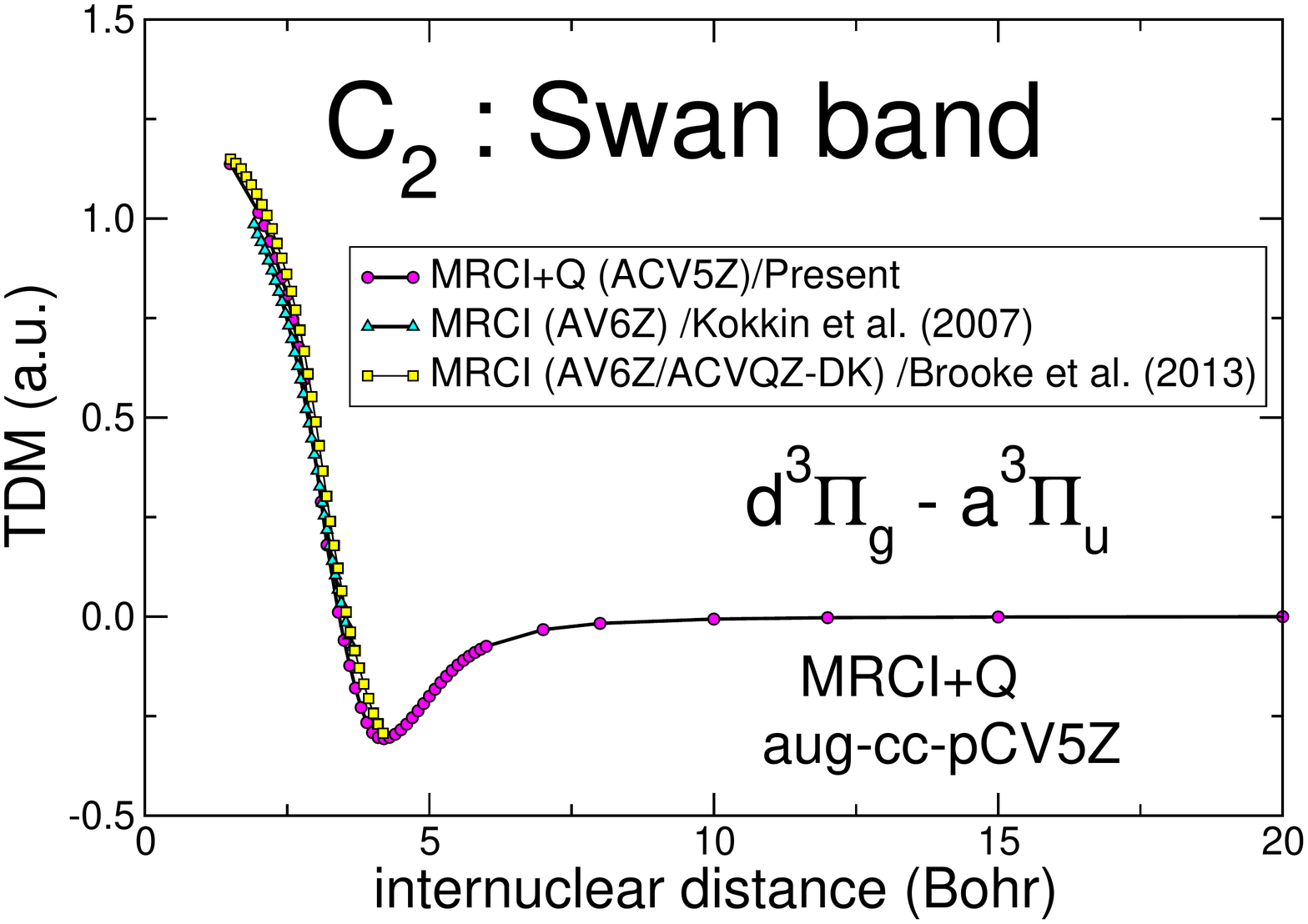}{8.9cm}{(a)}
             \fig{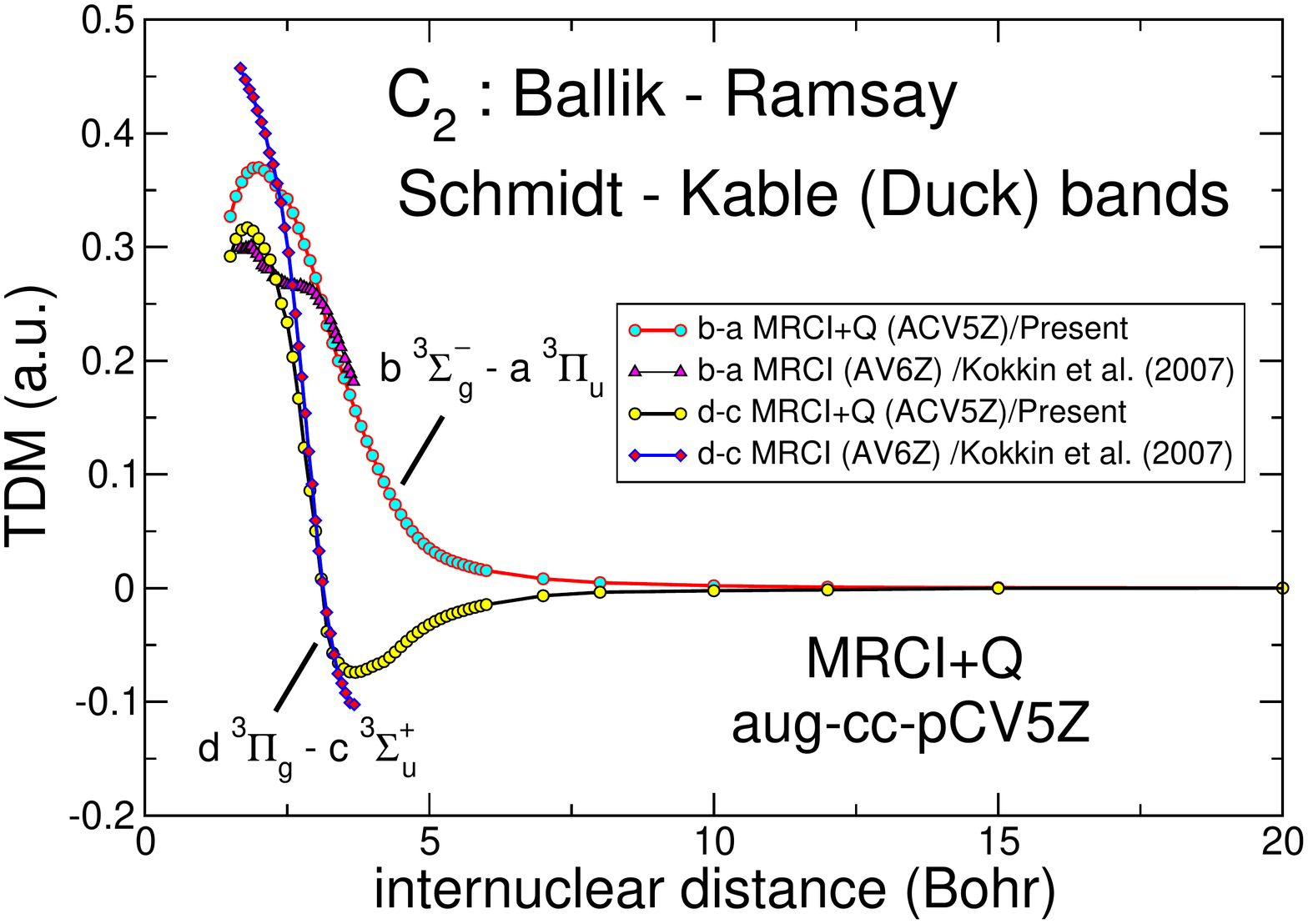}{8.9cm}{(b)}
             }
    \fig{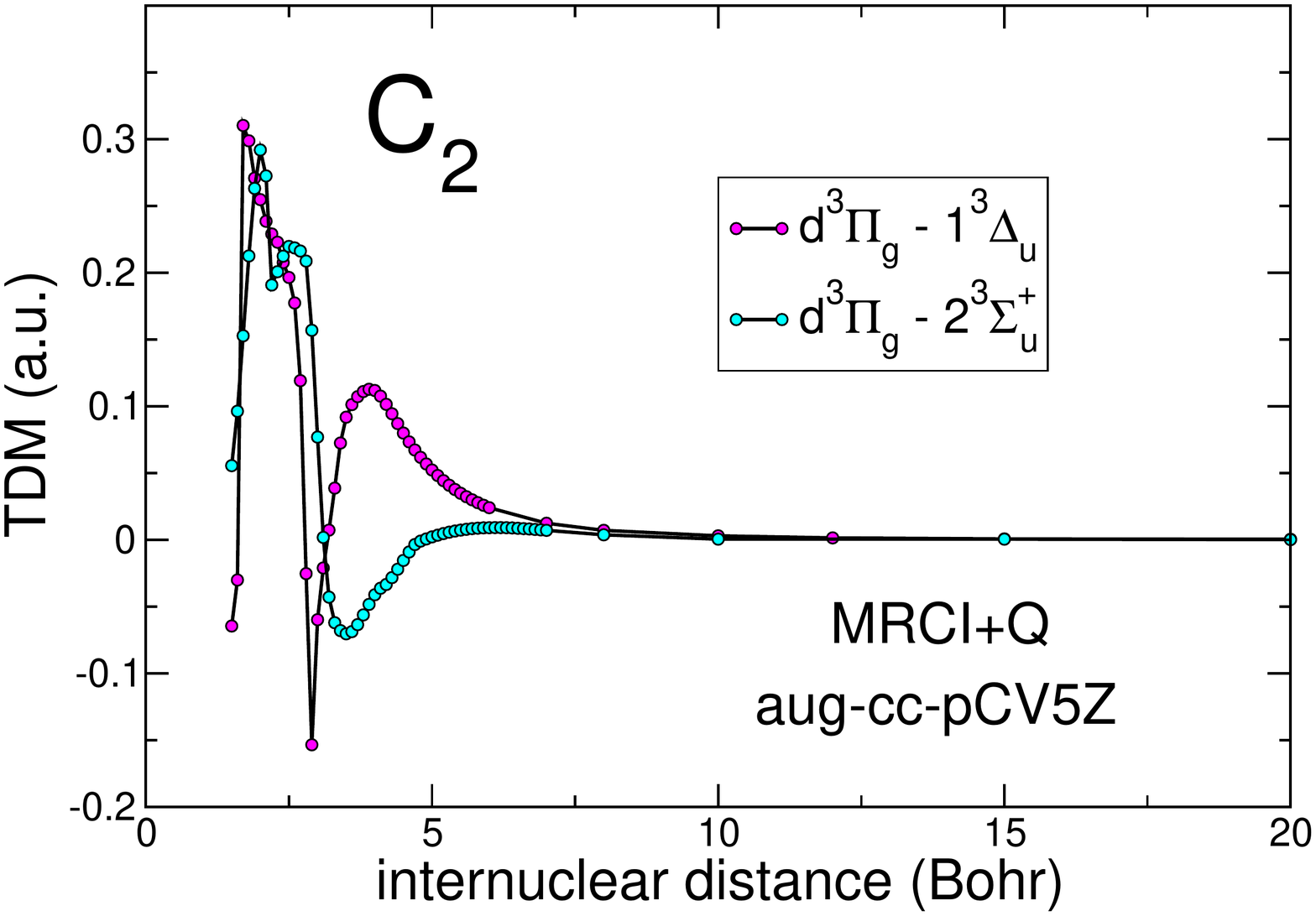}{8.9cm}{(c)}
    \caption{Transition dipole moments for triplet transitions in C$_2$ in atomic units, (a) Swan band (d$^3\Pi_g$-a$^3\Pi_u$), 
                   (b) Ballik-Ramsay (b$^3\Sigma_g^-$- a$^3\Pi_u$), and the Schmidt-Kable (Duck) bands (d$^3\Pi_g$-c$^3\Sigma_u^+$). 
                   Other triplet transition dipole moments are shown in (c) for the (d$^3\Pi_g$-1$^3\Delta_u$) 
                   and the (d$^3\Pi_g$-2$^3\Sigma^+_u$) bands.
                   We compare the present MRCI+Q work with results from \citet{kokkin07}
                   for the Swan and Ballik-Ramsay bands and from \citet{brooke13} for the Swan band.
                  }
    \label{fig:tdmtrip}
\end{figure}
	\begin{figure}
    \centering
    \fig{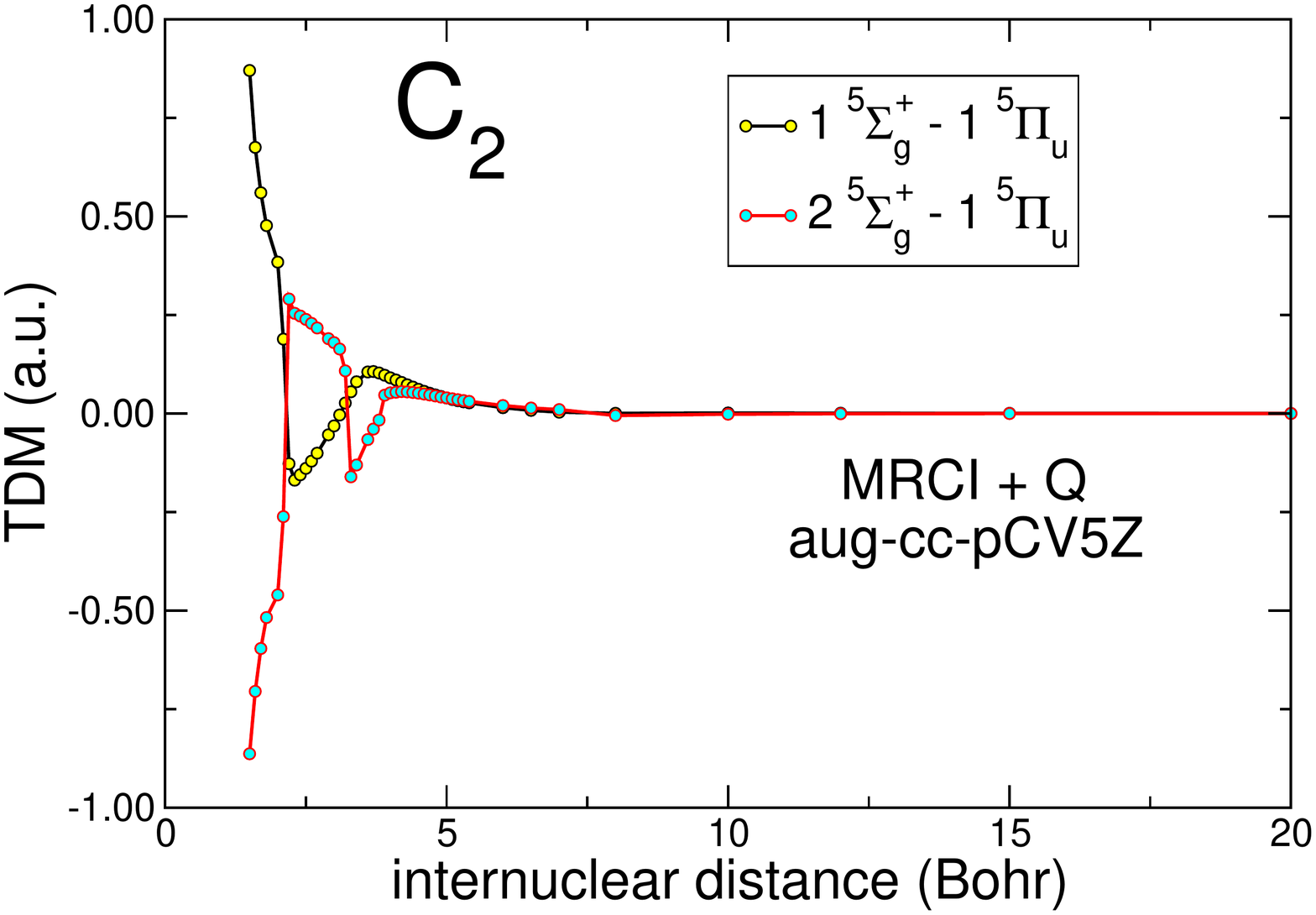}{8.9cm}{(a)}
    \fig{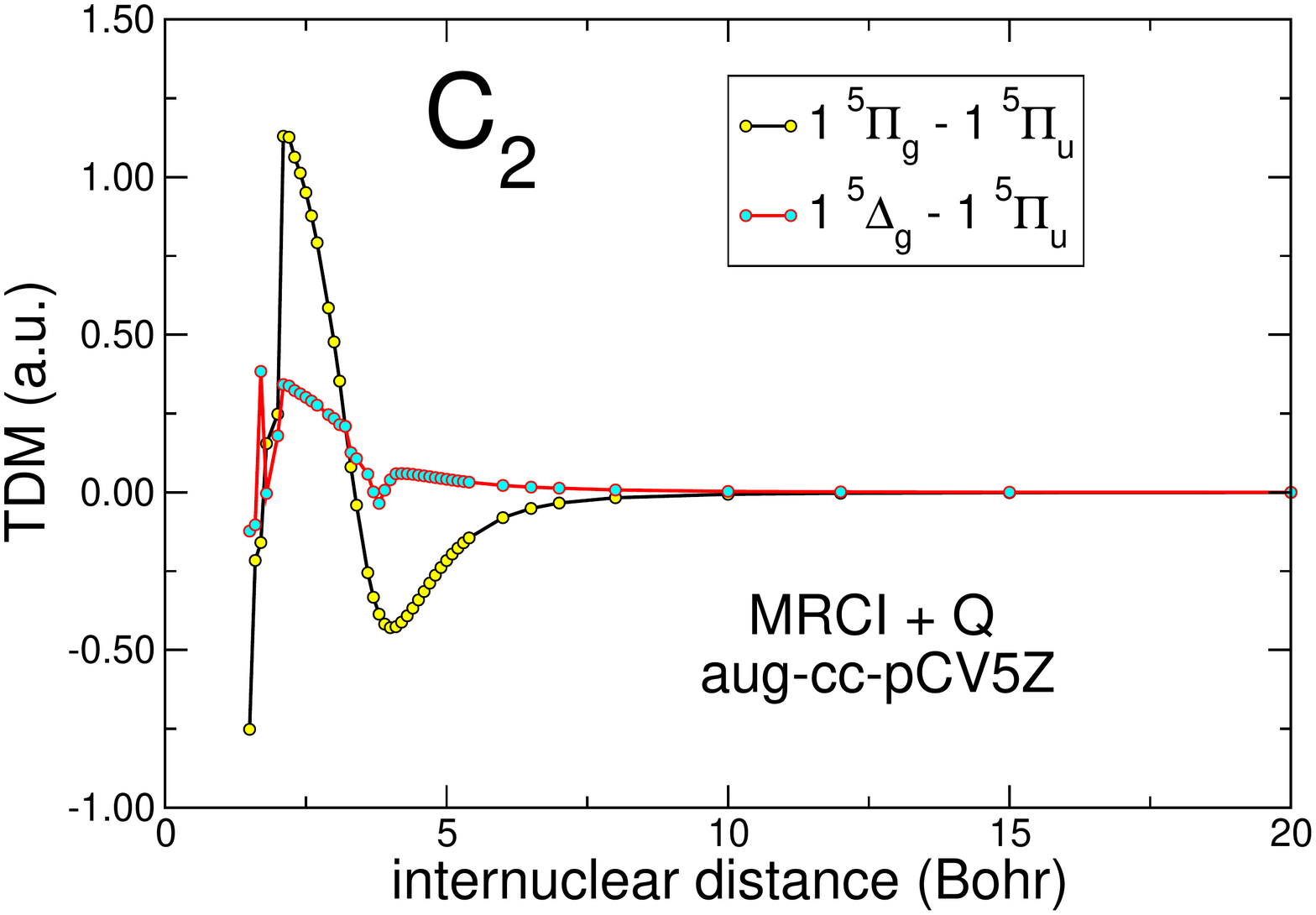}{8.9cm}{(b)}
    \caption{Transition dipole moments for quintet transitions in C$_2$ in atomic units, (a)  
                   (1$^5\Sigma_g^+$-1$^5\Pi_u$) 
                   and the (2$^5\Sigma_g^+$-1$^5\Pi_u$) bands, and in 
                   (b)  for the (1$^5\Pi_g$-1$^5\Pi_u$), and  (1$^5\Delta_g$-1$^5\Pi_u$) bands.
                   }
    \label{fig:tdmquint}
\end{figure}
The TDMs for singlet transitions are shown in Fig.~\ref{fig:tdmsing}, for triplet transitions in Fig.~\ref{fig:tdmtrip},
and for 
quintet transitions in Fig.~\ref{fig:tdmquint}.
As shown in Figs.~\ref{fig:tdmsing} and \ref{fig:tdmtrip} the results
are in satisfactory agreement with previous calculations for
the Phillips, Swan, Ballik-Ramsay, and Schmidt-Kable (Duck) bands
\citep{oneil87,langhoff90,kokkin07,brooke13}.
In addition, our results for the Bernath B$^{\prime}\Sigma_g^+$-A$^1\Pi_u$ and Deslandres-d'Azambuja bands
are provided over a substantially larger range of internuclear
distances compared to the earlier MRDCI calculations of \citet{chabalowski83}.
(The band 1$^5\Sigma_g^+$-1$^5\Pi_u$ has not yet been observed~\citep{bornhauser15}; we point
out that 2$^3\Sigma_u^+$-d$^3\Pi_g$,
1$^3\Delta_u$-d$^3\Pi_g$, and
1$^5\Delta_g$-1$^5\Pi_u$ bands may exist.
Additionally, we observe that the 2$^3\Sigma_u^+$-d$^3\Pi_g$ band
might contribute at wave numbers
where significant spectral congestion is seen in dicarbon~\citep{tanabashi07}.)

\section{Cross Sections}\label{sec:results}
The quantum mechanical cross section for the radiative association process 
$\sigma^{\textrm{QM}}_{i \rightarrow f}(E)$,
where the initial $i$ and final $f$ electronic states 
are labeled by their molecular states (e.g. d$^3\Pi_g$)
can be calculated using perturbation theory 
(see, for example, \citep{Babb1995,Franco1996}  and \citep{Babb1998}). The result is
\begin{equation}
 \sigma^{QM}_{i \rightarrow f} (E)  = P_{i}\sum_{v^{\prime}J^{\prime}}^{}\sum_{J}^{}
  \frac{64}{3} \frac{\pi^5}{137.036^3} \frac{\nu^3}{2\mu E} \mathcal{S}_{J J^{\prime}}
   |M_{i E J, f v^{\prime} J^{\prime}}|^2, \label{quantum}
\end{equation}
where the sum is over initial partial waves with 
angular momenta $J$ and final vibrational $v^{\prime}$
and rotational $J^{\prime}$ quantum numbers, 
$\mathcal{S}_{J,J^{\prime}}$ are the appropriate line strengths \citep{Cowan1981,Curtis2003} 
or H\"{o}nl-London factors \citep{Watson2008}, $137.036$ is the speed of light in atomic units, 
$\mu$ is the reduced mass of the collision system, and  
$M_{i E J, f v^{\prime} J^{\prime}}$ is given by the integral
\begin{equation}
{ M_{i E J, f v^{\prime} J^{\prime}}}
=\int_{0}^{\infty} F_{i E J}(R) D_{if}(R) 
\Phi_{f v^{\prime} J^{\prime}} (R)dR.
\label{matrix}
\end{equation} 
The wave function  $\Phi_{\Lambda^{\prime} v^{\prime} J^{\prime}} (R)$
is a bound state wave function of the final electronic state, 
$F_{\Lambda EJ} (R)$ is an energy-normalized continuum wave function
of the initial electronic state,
and $D_{if}(R)$ is an electric dipole transition dipole moment
between $i$ and $f$.

Due to presence of identical nuclei and the absence of
nuclear spin in $^{12}\mbox{C}_2$, the rotational quantum numbers
of the $^1\Sigma_g^+$ states are even  
and for any given value of $\Lambda=1$ or 2, only one lambda-doubling
level is populated~\citep{amiot83}.
Thus, the statistical weight factor $P_i$ is given by
\begin{equation}
\label{eq:P}
P_i = (2S_i+1)/81,
\end{equation}
where $S_i$ is the total spin of the initial molecular electronic state (here 1, 3, or 5), 
and there are for two C$(^3P)$ atoms $3^4=81$ molecular states labeled by $\Lambda$ and $S$.
Thus,
for example, for
the $^{12}\mbox{C}_2$ molecule considered here,
$P_i=\frac{1}{81}$, $\frac{3}{81}$ or $\frac{5}{81}$, respectively,
for $i=A^1\Pi_u$, $b^3\Sigma_g^+$, or $1^5\Pi_g$.

\begin{deluxetable}{c@{$~$}Lc@{$~$}Lc}[ht!]
	\centering
    \tablecaption{Transitions studied in this work. Listed in
    order of decreasing contribution to total cross section.\label{table:details}}
    \tablehead{
	\twocolhead{Initial} & 
        \twocolhead{Final} & 
        \colhead{Band Name} 
		}
    \startdata
    C&$^1\Pi_g$			& A&$^1\Pi_u$  	& Deslandres-d'Azambuja \\
    d&$^3\Pi_g$ 		& a&$^3\Pi_u$  	& Swan 					 \\
    1&$^5\Pi_u$ 		& 1&$^5\Pi_g$  	& Radi-Bornhauser 		   \\
    2&$^3\Sigma_u^+$ 	& d&$^3\Pi_g$  	& \nodata					\\
    1&$^3\Delta_u$ 		& d&$^3\Pi_g$  	& \nodata					\\
    b&$^3\Sigma_g^-$ 	& a&$^3\Pi_u$  	& Ballik-Ramsey 		 \\
    d&$^3\Pi_g$ 		& c&$^3\Sigma_u^+$ & Schmidt-Kable		  \\
    A&$^1\Pi_u$			& X&$^1\Sigma_g^+$ & Phillips 			\\
    B$^{\prime}$&$^1\Sigma_g^+$&A
    					&$^1\Pi_u$ 		& Bernath B$^\prime$ 		  \\
    B&$^1\Delta_g$ 		& A&$^1\Pi_u$ 	& Bernath B 			 \\
    2&$^5\Sigma_g^+$    & 1&$^5\Pi_u$  	&  	\nodata				  \\
    \enddata
\end{deluxetable}
The bound and continuum state wave functions may be computed from their
respective Schr\"odinger equations using the grid-based approach of the
Numerov method \citep{Cooley61,johnson77},
where we used step sizes of up to 0.001~Bohr. 
For example, for the $\text{X}\ ^1\Sigma_g^+$ state, we find 55 vibrational levels and 
$J=0,2,...,174$, though not all levels will contribute to the cross sections.
A finely spaced energy grid is required to account for  resonances. 
\begin{figure}
    \includegraphics[width=\textwidth]{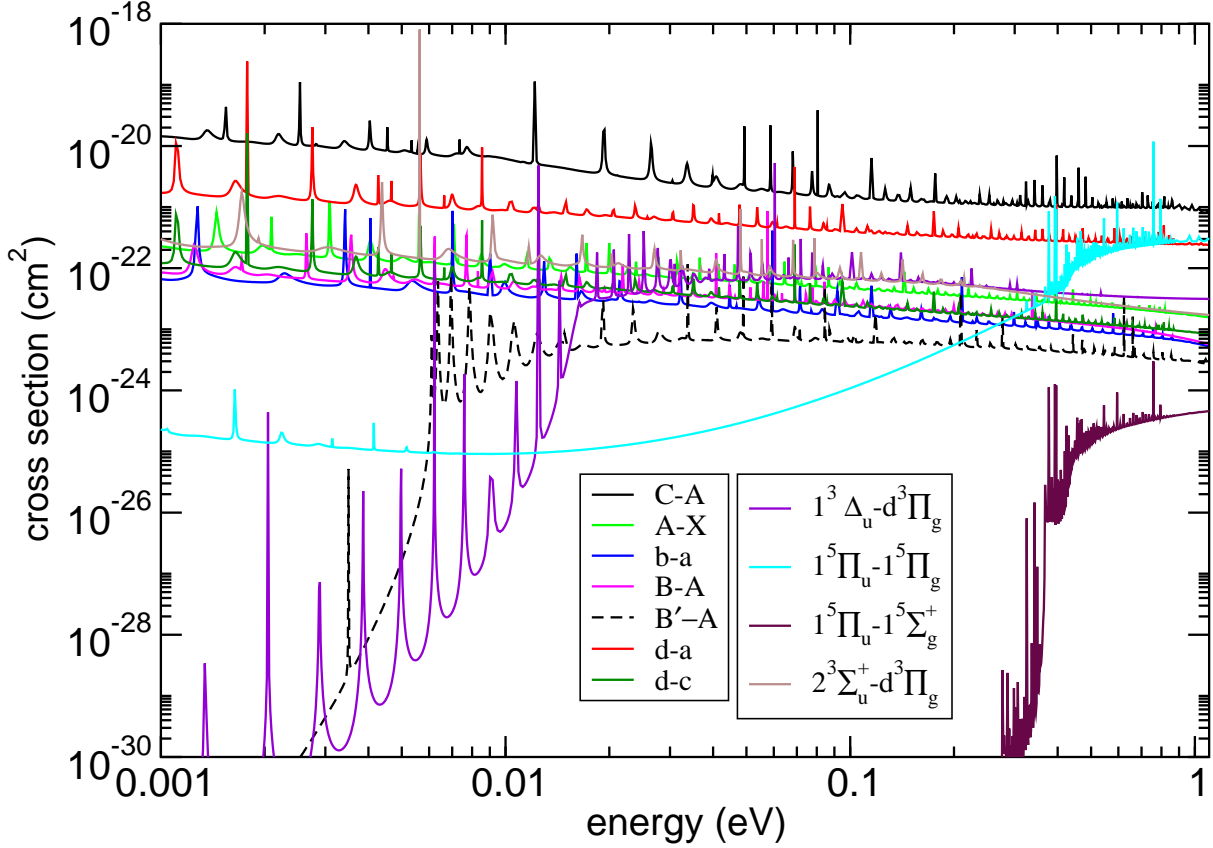}
    \caption{Radiative association cross sections (units of cm$^2$) 
    as a function of the collision energy $E$~(eV) for the collision of two $^{12}\mbox{C}(^3P)$ atoms.  
                   Results are shown for the singlet, triplet,
                   and quintet
                   transitions listed
                   in Table~\ref{table:details}.
                   In order to gauge the cross sections,
                   all data are plotted with the statistical factor $P_{i}=1$.  We see that the 
                   Deslandres-d'Azambuja (C$^1\Pi_g$-A$^1\Pi_u$), Swan (d$^3\Pi_g$-a$^3\Pi_u$)
                   and Radi-Bornhauser (1$^5\Pi_u$-1$^5\Pi_g$) bands have the largest cross sections.}
        \label{fig:cross-sections}
\end{figure}
In Fig.~\ref{fig:cross-sections}, results are shown for the radiative association cross section as a function 
of energy for the singlet, triplet, and quintet transitions listed in Table~\ref{table:details}.
We plot the cross sections with the statistical factor $P_i$ set equal to unity for all states.
Numerous  shape resonances are visible.
For the 1$^3\Delta_u$-d$^3\Pi_g$
and  1$^5\Pi_u$-1$^5\Pi_g$
cross sections, resonance tunneling features
are visible, corresponding to potential barriers (local maxima) 
in the entrance channels.
The local maxima from the potential energy curves
are
$0.015$~eV at $R=6$ for 1$^3\Delta_u$, 
$0.37$~eV at $R=3.6$ for 2$^5\Sigma_g^+$,
and 
$0.42$~eV at $R=3.5$ for 1$^5\Pi_u$ 
and as seen in Fig.~\ref{fig:cross-sections} 
the corresponding cross sections sharply diminish for collision energies
below these values.
Note, the cross sections for Deslandres-d'Azambuja transitions (C$^1\Pi_g$-A$^1\Pi_u$) dominate
the other cross sections for all collision energies,
followed by the cross sections for the Swan (d$^3\Pi_g$-a$^3\Pi_u$) transitions.
The cross sections for Radi-Bornhauser transitions (1$^5\Pi_u$-1$^5\Pi_g$) 
rise sharply as the relative energy increases.
The other cross sections are at least an order of magnitude weaker than
those corresponding to Deslandres-d'Azambuja, Swan, and Radi-Bornhauser transitions
and will not contribute significantly to the total cross section.
While the $1^5\Pi_u$ state does support
some bound levels~\citep{bornhauser17}, the calculated cross sections
from the 2$^5\Sigma_g^+$-$1^5\Pi_u$ state
are negligible because of the steeply repulsive tail of the initial 2$^5\Sigma_g^+$ electronic
state.

\subsection{Rate Constant}
The potential energy curves and transition dipole moments were then used to 
calculate the cross sections and rates for radiative association in the C$_2$ molecule.
The thermal rate constant (in $\mbox{cm}^3 \mbox{s}^{-1}$) at a given temperature $T$
to form a molecule by radiative association is given by
\begin{equation}
k_{i\rightarrow f} ={\left( \frac{8}{\mu\pi}\right )}^{1/2}
{\left( \frac{1}{k_B T}\right)}^{3/2} \int_{0}^{\infty} E\;  
\sigma_{i\rightarrow f} (E)\;   e^{-E/k_B T} dE\ ,
\label{rateconstant}
\end{equation}
where  $k_B$ is Boltzmann's constant.

The complicated resonance structures
make it challenging to calculate accurately the rate coefficient
using numerical integration \citep{Bennett2003,Gustafsson2012}. 
As a guide,
the relationship $v\sigma^\textrm{QM}_{i\rightarrow f}$, 
where $v=\sqrt{2E / \mu}$,
is  used to designate an effective
energy-dependent rate coefficient, 
where $R(E)$ for a transition from state $i$ to state $f$ is given by,
\begin{equation}	
R(E) = \sqrt{ 2E / \mu }\,  \sigma^\textrm{QM}_{i\rightarrow f} (E)  
\end{equation}
This form is often used to define a 
quasi-rate coefficient rather than one averaged over a Maxwellian distribution
and was utilized in ultra-cold collisional studies~\citep{krems10,bmcl14}.  
\begin{figure}
    \includegraphics[width=\textwidth]{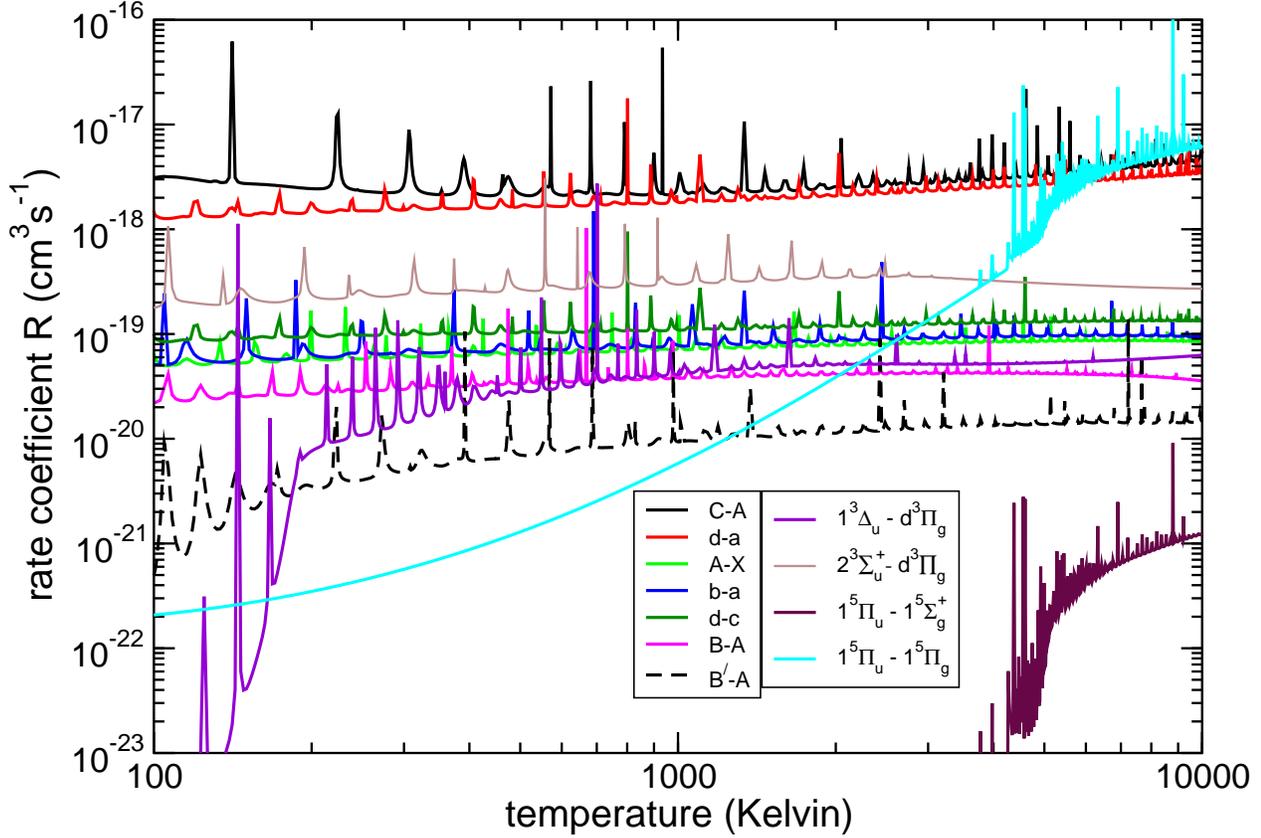}
    \caption{Radiative association quasi-energy dependent rates $R$ (cm$^3$/s) as a function of effective kinetic temperature (K) for the C$_2$ molecule.  
                   The appropriate weighting factors, Eq.~(\ref{eq:P}), are included.
                   The Deslandres-d'Azambuja (C$^1\Pi_g$-A$^1\Pi_u$),  
                    and 
                   the Swan (d$^3\Pi_g$-a$^3\Pi_u$) bands are the major contributors to the total rate, though
                   the Radi-Bornhauser (1$\,^5\Pi_u$-1$\,^5\Pi_g$) band contributes for $T> 5000$~K.}
        \label{fig:ratesv}
\end{figure}
In Fig.~\ref{fig:ratesv} results are shown for the radiative association rates $R(E)$ (in cm$^3$/s) as a function 
of energy expressed in temperature units (K) for the cross sections calculated for  the C$_2$ molecule. 
The Deslandres-d'Azambuja (C$^1\Pi_g$-A$^1\Pi_u$) transitions are the main
contributors for low temperature and the   
Radi-Bornhauser (1$\,^5\Pi_u$-1$\,^5\Pi_g$) bands
are the main contributors for high temperatures. 
The Swan (d$^3\Pi_g$-a$^3\Pi_u$) bands also contribute to the total rate.
\begin{figure}
    \includegraphics[width=\textwidth]{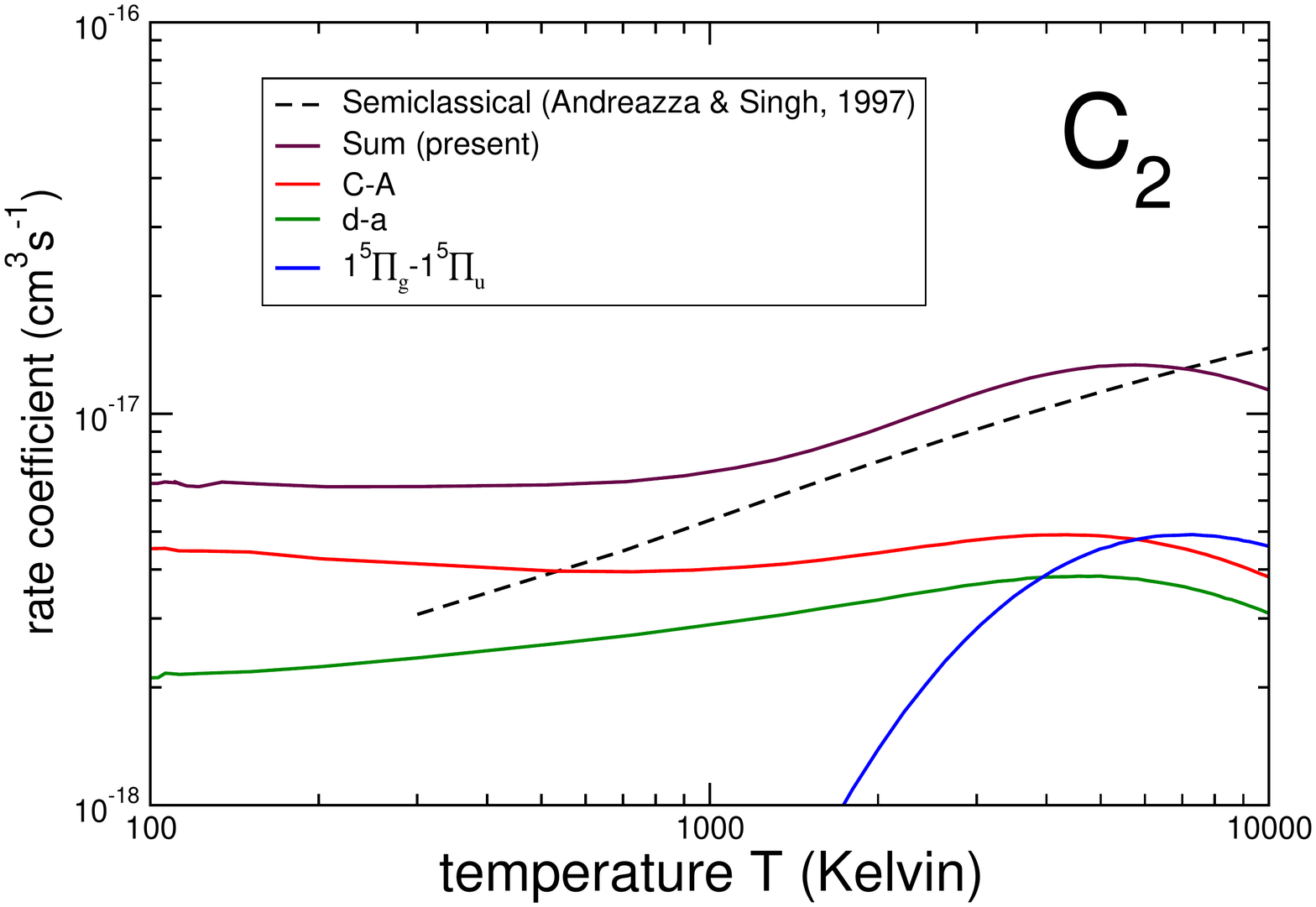}
    \caption{Maxwellian averaged radiative association rates (cm$^3$/s) as a function of temperature (Kelvin) for the C$_2$ molecule.  
                   Results are shown for the dominant singlet, triplet,
                   and quintet transitions with their appropriate statistical factor included.
                   The total quantal rate (brown line) is seen to lie  above the previous total 
                   semiclassical rate  \citep{andreazza97}  (dashed black line)
                   at all but the highest temperatures. }
        \label{fig:ratesm}
\end{figure}

In Fig.~\ref{fig:ratesm} we compare our Maxwellian averaged quantal  rates 
for the dominant Deslandres-d'Azambuja, Swan, and Radi-Bornhauser
transitions and their sum
with those determined from the previous 
semi-classical approximation by \cite{andreazza97}  over the temperature range 100-10,000 Kelvin.  
The quantal rates have the appropriate statistical population included 
so a comparison could be made directly with the previous semiclassical results of \citet{andreazza97}.
The total rate coefficient is fit to better than 6 percent  by the function
\begin{eqnarray}
\label{eq:fit}
\alpha(T) &=& 5.031 \times 10^{-18} + 1.501\times10^{-16} T^{-1}+ 2.517\times 10^{-21} T -1.89\times 10^{-25} T^2\quad \textrm{cm}^3/\textrm{s},\nonumber\\
& & \quad 100\leq T\leq 10,000~K
\end{eqnarray}
  
Contemporary discussions of resonances in radiative association cross sections
were given by, for example, \citet{Bennett2003,BarvanHem06,BenDicLei08,augustovicova12,mrugala13,golubev13},
though earlier researchers
also considered the problem in detail \citep{GiuSuzRou76,FloRou79,GraMosRou83},
with considerations of radiative and tunneling contributions
to resonance widths.
Our procedures for calculating the cross sections, e.g. Eq.~(\ref{quantum}),
and the corresponding rate coefficients, as shown in Fig.~\ref{fig:ratesm},
include shape and resonance tunneling resonances.
Procedures for the precision treatment of the effects of
radiative decay have been developed:  \textit{i)} excise certain resonances
from the cross sections, recalculate them
including the sum of the partial widths for radiative 
decay and tunneling, and then insert them back;
\textit{ii)} add the separately calculated resonance cross sections to the semi-classical cross sections;
or \textit{iii)} add the separately calculated resonance cross sections
to a  background smooth base line derived from the quantum cross sections.
\citet{franz11} and \citet{antipov13} examined radiative association
of carbon and oxygen to form CO ($\mbox{A}^1\Pi$-$\mbox{X}^1\Sigma^+$),
where there is a local maximum in the $\mbox{A}^1\Pi$ state of 0.079~eV (900~K effective
temperature).
Previous calculations 
of rate coefficients corresponding to
semi-classical cross sections (using different
molecular data) were available for comparison~\citep{dalgarno90,singh99}.
Using the quantum-mechanical
theory [our Eq.~(\ref{quantum})], using semi-classical methods, and with specific
treatment of resonances, \citet{franz11} and \citet{antipov13}
found 
that for the rate coefficients corresponding to  quantum cross sections 
with Eq.~(\ref{quantum}) \citep{franz11} or to hybrid calculations using
procedure \textit{ii)} (semi-classical theory 
combined with additive quantum treatment of
resonances) \citep{antipov13} that resonances
contribute below about 900~K.
However,  the rate coefficients
for the quantum calculation, Fig.~2 of \citep{franz11}
and those
for the hybrid calculations, Fig.~4a of \citep{antipov13},
do not deviate until the temperature is below 100~K.
Precision treatment of radiative decay for the numerous resonances in our cross sections
might yield enhanced values at lower temperatures, but given the application
envisioned, namely to calculate corresponding
rate coefficients for applications to the chemical models of supernovae ejecta,
where many chemical reactions enter and few reaction
rate coefficients are known precisely
and where the temperatures of interest are perhaps of order 1000-2000~K,
the present procedures are satisfactory.

\section{Discussion}\label{sec:discussion}
We see that our present quantal rates are larger than those from the previous semiclassical 
results of \citet{andreazza97} at temperatures below 7000~K, 
and our quantal rates persist with
a value of about $7\times 10^{-18}$~$\mbox{cm}^3\mbox{s}^{-1}$ as temperatures approach 100~K.
\citet{andreazza97} listed the Swan and Deslandres-d'Azambuja transitions
(in that order)
as the leading contributors to the radiative association process.
Our results indicate that the Deslandres-d'Azambuja transitions dominate
the Swan transitions for all temperatures.
\citet{andreazza97} used a potential energy function for the C$\;^1\Pi_g$ state with
a barrier of $0.002$~eV; they did not list
the cross sections, but in the semiclassical
method the cross sections for C$\;^1\Pi_g$-A$\;^1\Pi_u$ transitions would be negligible
for relative energies less than effective collisional temperature of the barrier, about 25~K,
so this would not affect the comparison at thermal temperatures.
We note that \citet{andreazza97} did not include quintet
states (for which no experimental data were available at that time)
and except for the trend towards lower temperatures,
our total rate coefficient is in reasonable agreement with their calculation.
We note that  
the 1$\;^5\Pi_u$-1$\;^5\Pi_g$ transition 
is the main contributor (as seen from Fig.~\ref{fig:ratesm}) to the rate coefficient for temperatures above 5000~K.

As mentioned above, in laboratory carbon vapors generated by laser radiation, the presence
of dicarbon is confirmed by, primarily, radiation from the Swan bands,
but also from the Deslandres-d'Azambuja bands \citep{savastenko11}.
In light emission
from laser-induced expansion of carbon vapor from a graphite rod,
\citet{monchicourt91}  found evidence of associative collisions of two ground state carbon atoms.
However, the operative source of dicarbon depends on factors
such as temperature and distance from the graphite substrate and there is
evidence that dicarbon is formed by dissociation from the graphite~\citep{iida94}
or recombination through three-body reactions~\citep{savastenko11}.
Recent modeling at certain densities of laser-induced plasmas using local thermal equilibrium and
equations of states indicates C$_2$ may form in plasmas of Si, N, or Ar and C 
at characteristic temperatures roughly less than 5000~K~\citep{shabanov15},
corresponding to the relatively low temperature region of the plasma~\citep{degiacomo17}.
That dicarbon is formed by a recombination process involving two carbon atoms in laser-induced plasma chemistry
was shown experimentally using carbon isotopes by~\citet{dong13}.

The dominance of the 1$\;^5\Pi_u$-1$\;^5\Pi_g$ transitions in our calculations is interesting
in light of theories of dicarbon formation in laser plasmas.
In particular, \citet{little87} theorized that the d$\;^3\Pi_g$ $(v=6)$ band of dicarbon is populated
by recombination collisions of atomic carbon,
possibly in the presence of a third body,
through the 1$\;^5\Pi_g$ state (cf. \citet{caubet94}). 
Subsequently, \citet{bornhauser11} experimentally verified
that the  metastable 1$\;^5\Pi_g$ state 
perturbs the d$\;^3\Pi_g$ state~\citep{bornhauser15};
population of the 1$\;^5\Pi_g$ state  leads to Swan band ($v=6$) fluorescence.
Subsequent experiments demonstrated 
the existence of the 1$\;^5\Pi_u$ state~\citep{bornhauser15}.
Depending on the densities and applicable chemistries of laser generated carbon
vapors, the radiative association
process 1$\;^5\Pi_u$-1$\;^5\Pi_g$  might contribute to the production of dicarbon in the 1$\;^5\Pi_g$ state
enhancing the mechanism of \citet{little87}.

For models of carbonaceous dust production in core-collapse supernovae, the present results
provide improved rates for an initial step in carbon chain production.
Our rates
may be compared to rates from a kinetic theory based nucleation model~\citep{lazzati16}, which
are a factor of $10^6$ larger, and can be applied to  molecular nucleation models~\citep{sluder18}
or chemical reaction network models~\citep{yu13,clayton18}.

\section{Conclusions}\label{sec:conclusions}

Accurate cross sections and rates for the formation
of dicarbon by the radiative association process were computed for transitions 
from several excited electronic states using new \textit{ab initio} potentials and transition dipole moment functions. 
Where calculated values exist in the literature, our TDM functions
are in good agreement with previous
results, 
but our calculations are presented over a more extensive range of internuclear distances.
We also present TDM functions
for the
B$^1\Delta_g$-A$^1\Pi_u$ 
d$^3\Pi_g$-1$^3\Delta_u$, d$^3\Pi_g$-2$^3\Sigma^+_u$,
1$\;^5\Pi_u$-1$\;^5\Pi_g$ (Radi-Bornhauser band),
$1\;^5\Sigma_g^+$-$1\;^5\Pi_u$,
$2\;^5\Sigma_g^+$-$1\;^5\Pi_u$,
and $1\;^5\Delta_g$-$1\;^5\Pi_u$ transitions,
substantially extending the available TDM data for dicarbon.
We found that
at the highest temperatures the quintet state Radi-Bornhauser ($1\;^5\Pi_u$-$1\;^5\Pi_g$) transitions are the
dominant channel for radiative association, followed by the Deslandres-d'Azambuja (C$^1\Pi_g$-A$^1\Pi_u$) and Swan 
(d$^3\Pi_g$-a$^3\Pi_u$) transitions,
while at lower temperatures the Deslandres-d'Azambuja transitions are dominant.
The computed cross sections and rates for C$_2$ are suitable for applicability in a variety of interstellar 
environments including diffuse and translucent clouds and ejecta of core-collapse supernovae.
In addition, our calculations do not contradict evidence that in laser ablated vapors dicarbon
formation proceeds through the $1\;^5\Pi_g$ state.
We concur with \citet{furtenbacher16} that further experimental studies on the 
Deslandres-d'Azumbuja transitions are desirable.

\acknowledgements
This work was supported by a Smithsonian Scholarly Studies grant.
BMMcL acknowledges support by the US National Science Foundation through 
a grant to ITAMP at the Center for Astrophysics \textbar \ Harvard \& Smithsonian  under
the visitor's program, the University of Georgia at Athens for the award of an adjunct professorship, 
and Queen's University Belfast for a visiting research fellowship (VRF). 
We thank Captain Thomas J. Lavery, USN, Ret., for his constructive comments that 
enhanced the quality of this manuscript.
The authors acknowledge this research used grants of computing time at the National
Energy Research Scientific Computing Centre (NERSC), which is supported
by the Office of Science of the U.S. Department of Energy
(DOE) under Contract No. DE-AC02-05CH11231.
The authors gratefully acknowledge the Gauss Centre for 
Supercomputing e.V. (www.gauss-center.eu) 
for funding this project by providing computing time on the GCS Supercomputer 
HAZEL HEN at H\"{o}chstleistungsrechenzentrum Stuttgart (www.hlrs.de).
ITAMP is supported in part by NSF Grant No.\ PHY-1607396.

\label{lastpage}
\end{document}